\documentclass{ecai} 

\usepackage{latexsym}
\usepackage{amssymb}
\usepackage{amsmath}
\usepackage{amsthm}
\usepackage{booktabs}
\usepackage{enumitem}
\usepackage{graphicx}
\usepackage{color}
\usepackage{appendix}
\usepackage{float}
\usepackage{hyperref}
\usepackage{natbib}
\usepackage{xcolor}

\newcommand{\BibTeX}{B\kern-.05em{\sc i\kern-.025em b}\kern-.08em\TeX}

\begin{document}


\begin{frontmatter}


\paperid{123} 


\title{Can News Predict the Direction of Oil Price Volatility? A Language Model Approach with SHAP Explanations}


\author[A]{\fnms{Romina }~\snm{Hashami}\orcid{0009-0005-0953-8821}}
\author[A]{\fnms{Felipe}~\snm{Maldonado}\orcid{0000-0001-8272-5192}\thanks{Corresponding Author. Email: felipe.maldonado@essex.ac.uk}}

\address[A]{School of Mathematics, Statistics and Actuarial Science, University of Essex}


\begin{abstract}
Financial markets can be highly sensitive to news, investor sentiment and economic indicators, leading to important asset price fluctuations.  In this study we focus on crude oil, due to its crucial role in commodity markets and global economy. Specifically we are interested on understanding the directional changes of oil price volatility, and for this purpose, we investigate whether news alone -- without incorporating traditional market data -- can effectively predict the direction of oil price movements. Using a decade-long dataset from Eikon (2014–2024), we develop an ensemble learning framework to extract predictive signals from financial news. Our approach leverages diverse sentiment analysis techniques and cutting-edge language models, including FastText, FinBERT, Gemini, and LLaMA, to capture market sentiment and textual patterns. We benchmark our model against the Heterogeneous Autoregressive (HAR) model and assess statistical significance using the McNemar test. Notably, while most sentiment-based indicators do not consistently outperform HAR, the raw news count emerges as a robust predictor. Among embedding techniques, FastText proves most effective for forecasting directional movements. Furthermore, SHAP-based interpretation at the word level reveals evolving predictive drivers across market regimes: pre-pandemic emphasis on supply-demand and economic terms; early pandemic focus on uncertainty and macroeconomic instability; post-shock attention to long-term recovery indicators; and war-period sensitivity to geopolitical and regional oil market disruptions. These findings highlight the predictive power of news-driven features and the value of explainable NLP in financial forecasting.

\end{abstract}

\end{frontmatter}


\section{Introduction}

Crude oil is one of the most vital commodities in the global economy, serving as a primary energy source and a key driver of financial markets. West Texas Intermediate (WTI) and Brent crude are the two most influential oil benchmarks, with WTI sourced from North America and Brent from the North Sea \cite{deng2019prediction}. Because oil is strategically important, changes in its price have a big impact on financial markets, business decision-making, and government policies \cite{xu2010forecasting, charles2009efficiency}. For this reason, volatility forecasting is an essential task. Various stakeholders -- including policymakers, energy firms, hedge funds, and individual investors -- closely monitor oil price movements, particularly volatility and directional changes, as these influence risk management and investment decisions \cite{deng2019prediction}. Significant benefits would result from a precise forecast of the direction of crude oil price volatility, including better trading techniques, better company financial planning, and informed government policies \cite{he2022novel}.

Traditional econometric models, including the autoregressive integrated moving average (ARIMA) model, heterogeneous autoregressive (HAR) framework, and generalized autoregressive conditional heteroscedasticity (GARCH) models, have been widely used in crude oil price forecasting \cite{corsi2009simple, wickham1996volatility, wang2016forecasting}. Among these, the HAR model has been particularly effective in capturing volatility persistence over different time horizons (short, medium, and long-term). However, a key limitation of HAR and similar models is their exclusive reliance on historical price data, which may not fully capture external shocks that drive oil price volatility \cite{gkillas2020forecasting, gong2018incremental}. 

To address those limitations, is that researchers have explored alternative methods, such as the use of  Natural Language Processing (NLP), which has shown promising applications to finance, extracting valuable insights from textual data and news \cite{hu2021crude, lu2020crude}. Particularly, sentiment analysis has been utilized to gauge market movements,  using handpicked word lists and suitable machine learning techniques \cite{chen2021investor, bai2022crude}. With the advent of deep learning and Transformer-based models such as BERT and FinBERT, sentiment analysis has significantly improved, capturing more intricate linguistic patterns and contextual nuances \cite{gkillas2020forecasting}. Additionally, the use of Large Language Models (LLMs) has allowed to move beyond sentiment analysis, and being able to extract more detailed semantic information, transforming text into suitable numerical representations or  embeddings (e.g., \cite{touvron2023llama}). While these recent advancements show considerable promise, they are not without their shortcomings. Being one of them the lack of interpretability of their results, in contrast to econometric models. 


Naturally, the combination of desirable properties coming from econometric models, into text-based models presents an opportunity for improvement, although it is unclear how to systematically do this (since their inputs are drastically different), and how to asses their overall performance. As a starting point, it is desirable to recover some of the interpretability that econometric models might offer. And to achieve this we require that text-based models are coupled with additional tools, such as explainability techniques like SHAP (SHapley Additive exPlanations) \cite{lundberg2017unified}. Helping to identify which words and phrases most significantly influence predictions -- making these models more interpretable for financial decision-making. SHAP has also being used by \cite{lu2022oil} to understand the contribution of features for volatility prediction, but on top of machine learning models using time-series data (i.e. not textual data).


In our study, we investigate the impact of different types of textual features -- specifically sentiment scores and embeddings -- compared against HAR models, for  volatility forecasting. The embeddings are generated using both deep learning models (such as FinBERT and Gemini) and traditional approaches (including GloVe and FinText). Additionally, we adopt classification-based models instead of traditional regression methods, as classification is better suited for capturing nonlinear patterns, handling small fluctuations, and directly predicting the direction of volatility \cite{he2022novel}.


This study makes four key contributions:

1. {\it News-Based Volatility Forecast}: Unlike standard models based on past market information, our method uses only news data as input for predicting crude oil price volatility, showing the value of textual information as a sole predictor.

2. {\it Sentiment vs. Embedding-Based Analysis}: We comprehensively compare sentiment scores and embedding-based textual representations derived from LLMs to assess which method more effectively captures market-relevant information.

3. {\it Benchmarking Against HAR Model}: Our proposed news-driven classification model is evaluated against the HAR model, a widely used benchmark in volatility forecasting, to highlight the advantages of incorporating textual data over conventional econometric approaches.

4. {\it Interpretable Machine Learning for Textual Models}: To show the predictive mechanisms of our textual models, we apply SHAP values at the word level, offering a granular interpretation of how specific terms influence model predictions. 

This approach builds on and extends recent work by Parvini and Assa \cite{parvini2025textual}, providing novel empirical insights into the key textual drivers of volatility forecasts and advancing the interpretability of deep learning-based financial NLP models. Crude oil is used as a case study to illustrate both the explainability of our modeling technique and the forecasting usefulness of news material, however, the procedure can be easily applied to other areas. For reproducibility purposes, we provide the repository \url{https://github.com/Romina-Hashami/Textual_Direction_Prediction_Oil_Volatility}.


\section{Related works}
Crude oil price forecasting has traditionally relied on econometric models, such as autoregressive integrated moving average (ARIMA), generalized autoregressive conditional heteroskedasticity (GARCH), and their extensions, to capture volatility fluctuations \cite{herrera:2018}, \cite{kang:2009}. These models are widely adopted due to their ability to incorporate volatility persistence and structural breaks. For instance, the CGARCH and FIGARCH models have been shown to outperform standard GARCH models in forecasting crude oil volatility across various markets \cite{kang:2009}. More recent studies have expanded these frameworks by integrating external uncertainty indices. Zhang \cite{zhang:2024} introduces the energy-related uncertainty index (EUI) into the GARCH-MIDAS framework, demonstrating its superior forecasting capabilities. Similarly, Song \cite{song:2022} finds that petroleum-focused volatility trackers based on newspaper data (PEMV) significantly enhance oil market volatility predictions when compared to traditional financial volatility measures, such as the VIX.

Another widely used econometric model for volatility forecasting is heterogeneous autoregressivesive (HAR) model, first introduced by Corsi \cite{corsi2009simple}. The HAR model captures volatility persistence by incorporating realized volatility across multiple time horizons. Corsi and Reno \cite{corsi2009har} extended this framework with the HAR-J model, which accounts for jumps in volatility and heterogeneous leverage effects, making it a robust tool for analyzing volatility dynamics. Zhang \cite{zhang:2019} further advanced the HAR framework by comparing forecast combination approaches with shrinkage methods, such as elastic net and lasso. Their results demonstrated that these methods significantly outperform individual HAR models and forecast combination techniques in out-of-sample oil price volatility prediction.

Despite the growing relevance of carbon and oil price prediction in the context of climate change, modeling the nonlinear effects in time series remains a significant challenge. To address this Molina-Muñoz et al. \cite{molina2024predicting} propose a hybrid forecasting framework that combines random forest, support vector machines, autoregressive integrated moving average, and nonlinear autoregressive neural network models. Their study demonstrates that hybrid models, particularly those that integrate machine learning and traditional econometric techniques, significantly improve prediction accuracy by capturing both linear and nonlinear dynamics in carbon and oil price returns. This highlights the increasing trend of leveraging hybrid models that combine machine learning with conventional econometric approaches to enhance forecasting performance.


Machine learning (ML) approaches have gained traction due to their ability to model intricate relationships within high-dimensional data. Chen et al.\cite{chen:2021} emphasizes the role of investor sentiment and financial uncertainty indices in improving oil volatility forecasts. Similarly, Diaz et al. \cite{diaz:2024} compares ML-based models with the HAR framework, demonstrating that ensemble learning and deep learning models outperform traditional econometric approaches in oil price forecasting.

Most research so far has focused on using regression models to predict oil prices (e.g., \cite{mohsin2023novel}), but the application of  classification methods for this problem, remains underexplored. Lately, predicting which direction crude oil prices will move has become a growing area of interest, and machine learning is starting to show a lot of promise in this space. He et al. \cite{he2022novel} propose a hybrid forecasting model that combines variational mode decomposition and symbolic time series analysis to extract multi-modal data features. Their use of machine learning classifiers, particularly Support Vector Machines, shows superior performance in predicting oil price trends, especially during high-volatility periods. This underscores the potential of machine learning to enhance directional forecasting in oil price prediction. Al-Fattah \cite{al-fattah:2019} further contributes to this body of research by introducing a hybrid artificial intelligence approach that combines genetic algorithms, artificial neural networks, and data mining techniques to model and forecast oil price volatility. Their model, which is designed to forecast West Texas Intermediate (WTI) crude oil price volatility, achieved 88\% accuracy in predicting volatility movements.
This model not only captures the complex dynamics of oil price swings, but also provides useful applications for producers, consumers and investors who seek to mitigate market instabilities and enhance investment strategies.

\section{Methodology}
In this section, we address the task of predicting Brent crude oil price volatility directions using news data. We evaluate several types of textual features in separate experiments -- sentiment scores extracted from news articles, the number of news items, and embeddings from various LLMs. Each feature set is used independently to train classification models, enabling a focused comparison of their predictive performance against a benchmark HAR model.

The HAR model and other conventional econometric models only use historical market data.  Despite their effectiveness, they could ignore dynamic market dynamics and outside shocks. To address this issue, we look for additional sources of information, such as financial text that can now be used to extract predictive data thanks to developments in Natural Language Processing. 

Previous studies have used sentiment indices from financial news \cite{araci2019finbert, sinha2022sentfin}, news volume as a proxy for media attention \cite{engle2020hedging}, and pre-trained models such as FinBERT to analyze financial text \cite{huang2023finbert}. However, reducing rich textual data to sentiment scores may lead to the loss of important information. To address this, we also employ Transformer-based embeddings that retain the high-dimensional structure of textual content. We compare these embeddings against sentiment-based features to assess their predictive power for oil price volatility direction. All models are benchmarked against the HAR model using a rolling window approach for out-of-sample performance evaluation.

\begin{figure}[h]
\centering
\includegraphics[width=7.5cm]{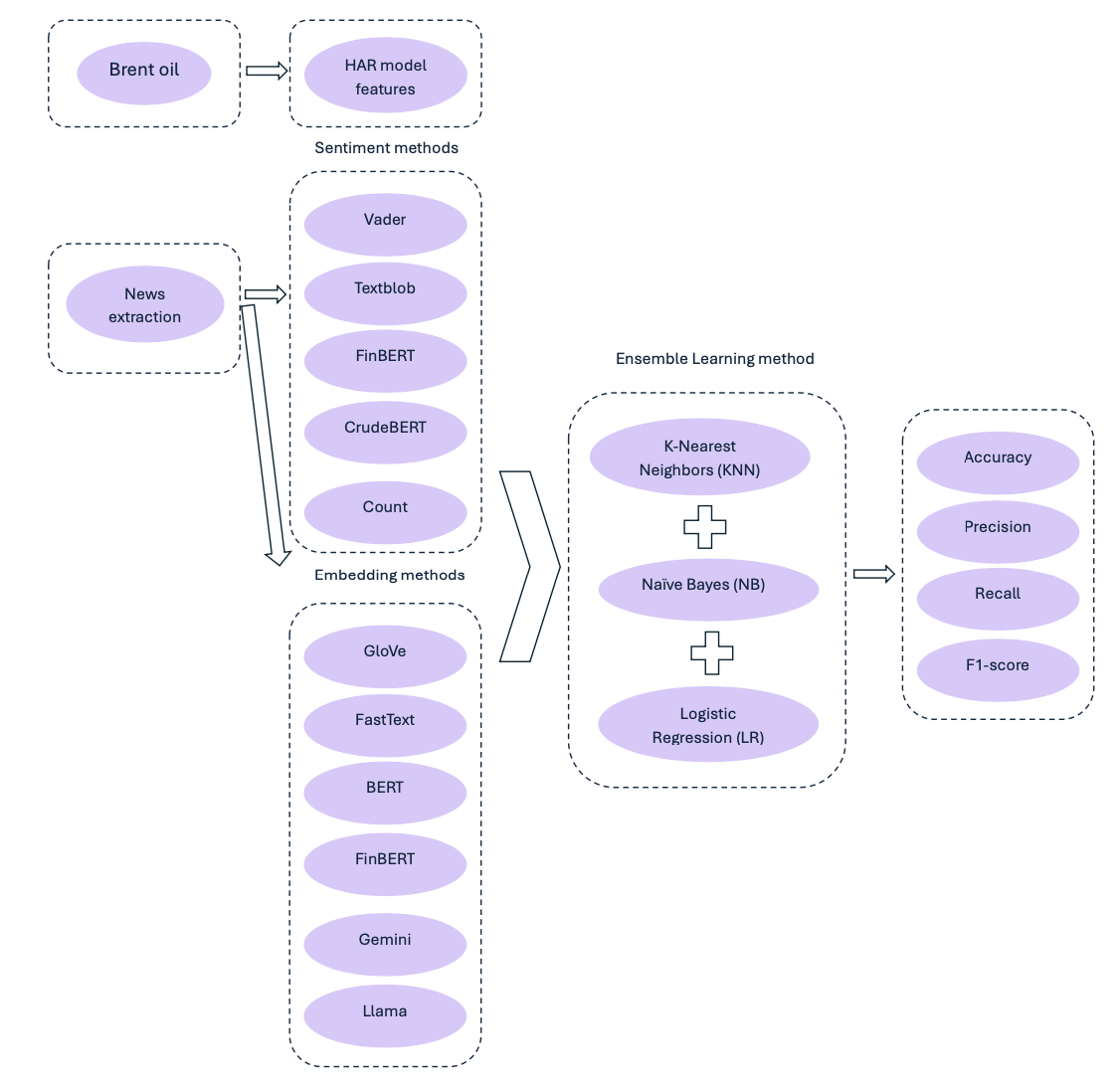}
\caption{Workflow.}
\label{fig:Flowchart_methodology}
\end{figure}

This methodology bridges traditional econometric models and NLP-based forecasting, demonstrating the value of language models in financial prediction. Figure \ref{fig:Flowchart_methodology}  illustrates the complete workflow used in this study. 

\subsection{Data collection and processing}
For this study, we collected news articles from Thomson Reuters (Eikon) to analyze factors influencing crude oil market realized volatility, covering Brent crude oil. The dataset consists of 592,858 news headlines spanning from January 2014 to December 2023. Headlines were selected for their capacity to offer concise descriptions of significant occurrences, which makes them appropriate for effective analysis \cite{huang2023finbert}.  We used regular expressions (RegEx) to clean the dataset by eliminating dates, email addresses, URLs, boilerplate content, and paragraphs that were too short. 

\begin{figure}[h]
\centering
\includegraphics[width=7.5cm]{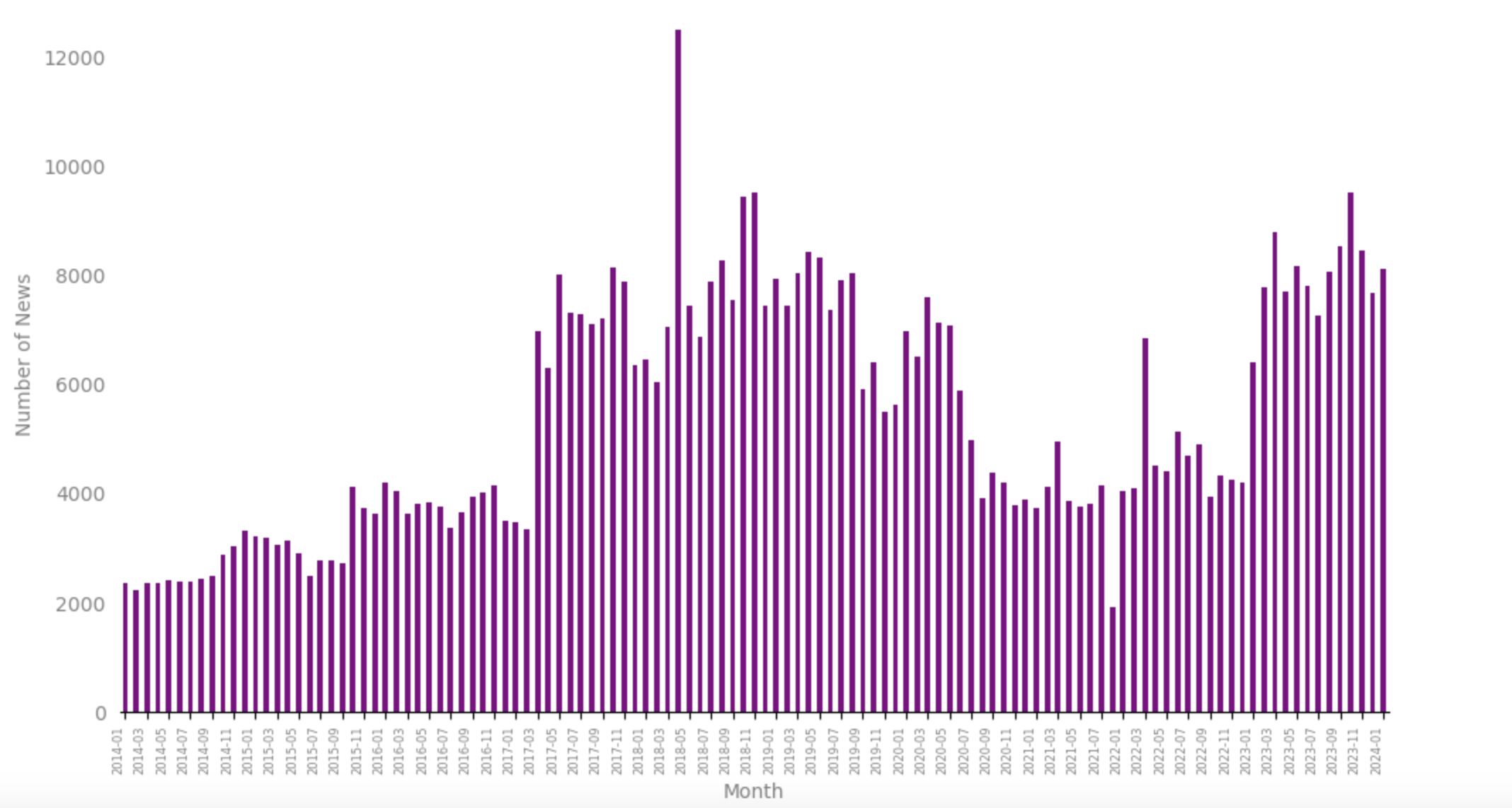}
\caption{Monthly distribution of news articles.}
\label{fig:news_volume}
\end{figure}

Table~\ref{tab:token_outputs_clean} shows the tokenized outputs generated by different embedding models for the input headline. It also reports the dimensionality of the resulting feature vectors, indicating how many features each model uses to represent the input text. Figure~\ref{fig:news_volume} presents the monthly volume of collected news articles, highlighting spikes during major events such as OPEC meetings, OPEC+ ministerial sessions, and global crises like the COVID-19 pandemic. 

\begin{table*}[h]
\caption{Tokenized outputs from some embedding models for the headline: 
\textit{``New Trump administration plan could boost oil drilling on remote Alaska reserve''} 
(Date: 2019-11-22)}
\vspace{1em}
\label{tab:token_outputs_clean}
\centering
\begin{tabular}{l@{\hspace{6mm}}c@{\hspace{6mm}}l}
\toprule
\textbf{Tokenizer} & \textbf{Vector Dim.} & \textbf{Tokens} \\
\midrule
FastText & 300 & ['New', 'Trump', 'administration', 'plan', 'could', 'boost', 'oil', 'drilling', 'on', 'remote', 'Alaska', 'reserve'] \\
FinBERT & 768 & ['New', 'Trump', 'admin', 'istration', 'plan', 'could', 'boost', 'oil', 'drill', 'ing', 'on', 'remote', 'Alaska', 'reserve'] \\
Gemini Pro & 768 & [' New', ' Trump', ' administration', ' plan', ' could', ' boost', ' oil', ' drilling', ' on', ' remote', ' Alaska', ' reserve'] \\
\bottomrule
\end{tabular}
\end{table*}

\begin{figure}[h]
\centering
\includegraphics[width=7.5cm]{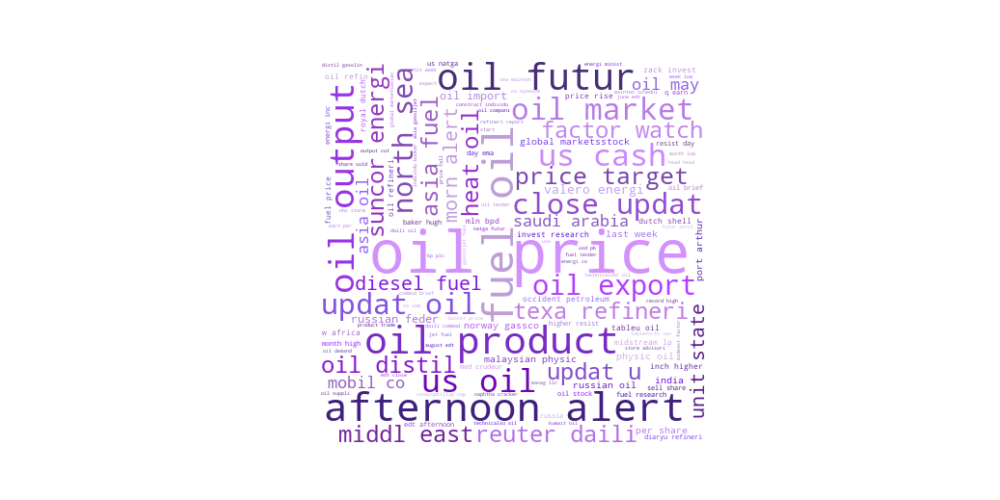}
\caption{Word clouds for news headlines}
\label{fig:wordcloud}
\vspace{1.5em}
\end{figure}

Figure~\ref{fig:wordcloud} further illustrates the most frequently occurring terms in the dataset through a word cloud. Larger font sizes represent terms that appear more frequently, emphasizing dominant topics in the dataset. The dataset's primary focus on price fluctuations, production levels, and trade flows is shown by key terms like \textit{"oil price," "fuel," "output," "export," "refinery,"} and \textit{"market"}.  Geographical allusions like \textit{"Russia," "Texas," "Asia," "OPEC,"} and \textit{"North Sea"} also draw attention to important areas that affect the oil market. The visualization confirms that the dataset is rich in domain-specific information, making it highly relevant for sentiment analysis and volatility prediction.

\vspace{1em}

\begin{figure}[h]
\centering
\includegraphics[width=7.5cm]{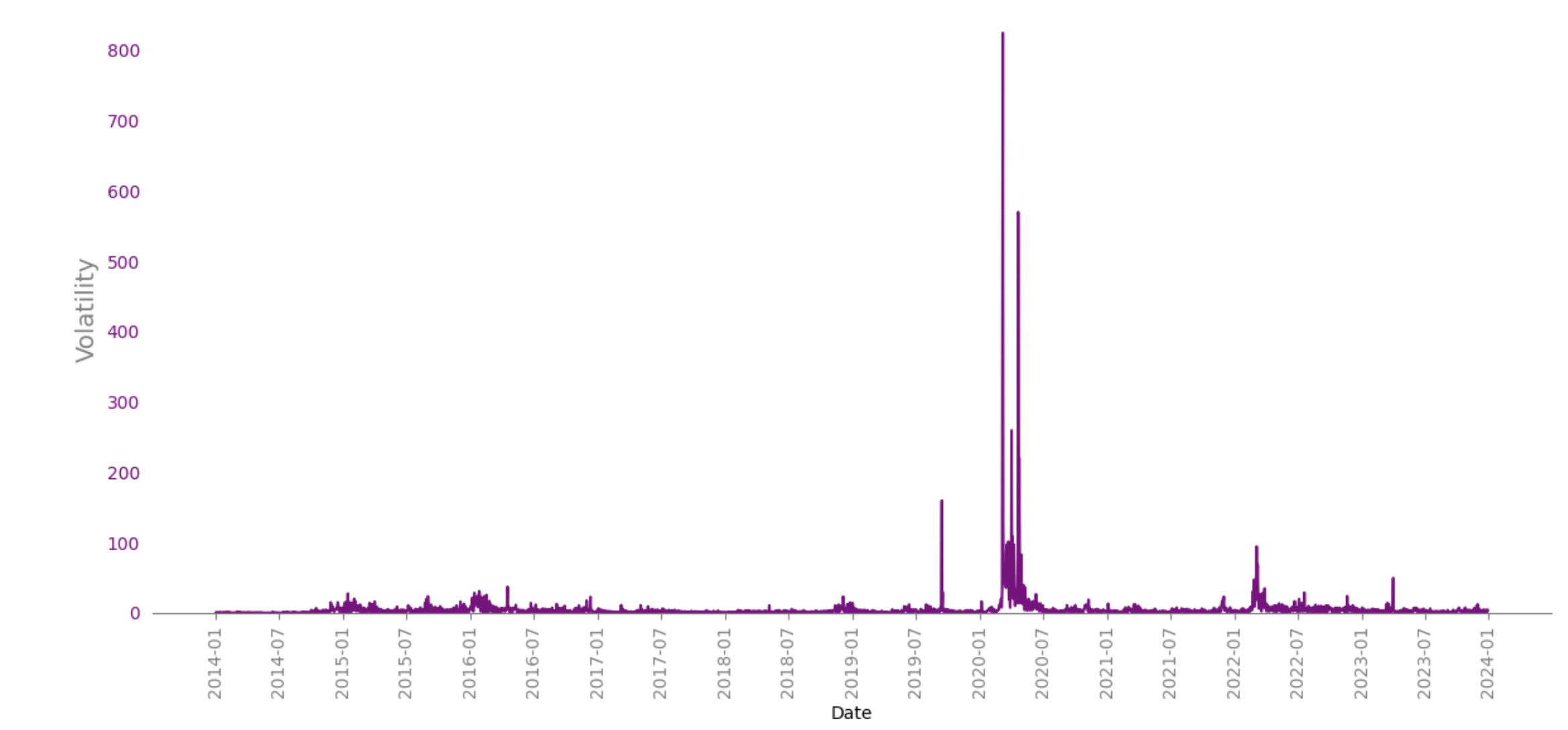}
\caption{Brent oil price volatility.}
\label{fig:eurai}
\end{figure}

We gathered historical high-frequency data on Brent oil futures from the \url{https://www.barchart.com} website to analyze market dynamics and capture intraday price fluctuations. The realized volatility of Brent oil prices was computed using 5-minute intraday log returns. Figure~\ref{fig:eurai} illustrates the realized volatility from 2014 to 2024, highlighting notable fluctuations associated with major events and market disruptions.

In the realized volatility literature, the log-price process of an asset $P_t$ is typically represented as an It\^o process ~\citep{barndorff2002econometric,andersen2003modeling}:
\[
d\log(P_t) = \mu_t\,dt + \sigma_t\,dW_t,
\]
where $\mu_t$ is the drift term, $\sigma_t$ is the instantaneous volatility, and $W_t$ is a standard Brownian motion. Within this framework, the daily integrated variance over the interval $[t-1, t]$ is defined as
\[
IV_t = \int_{t-1}^{t} \sigma_s^2\,ds.
\]

Since the integrated variance is not directly observable, it can be consistently estimated by the realized variance (RV), computed as the sum of squared intraday returns~\citep{andersen1998answering,barndorff2002econometric,jacod2003limit}:
\[
RV_t = \sum_{i=1}^{n} r_{t,i}^2,
\]
where the intraday return for the $i$-th interval on day $t$ is
\[
r_{t,i} = \log \left( P_{t-1+i\Delta} \right) - \log \left( P_{t-1+(i-1)\Delta} \right),
\]
and the sampling frequency is $M = 1/\Delta$. As $\Delta \to 0$, $RV_t$ converges in probability to $IV_t$~\citep{barndorff2002econometric,jacod2003limit}.

In practice, however, very high-frequency returns are contaminated by market microstructure noise. To mitigate this, the literature has established the use of 5-minute returns as a practical standard, balancing efficiency and noise reduction~\citep{andersen2000great,LiXiu2016}.

While realized volatility is frequently employed in volatility forecasting models~\citep{bollerslev2016exploiting}, in this study it is used as a model-free, ex-post estimate of the actual price variability in Brent oil futures markets.

\vspace{1.5em}

The data shows distinct spikes in volatility aligned with major global events. For instance, the Brent data clearly shows the oil price fall of 2014–2016, which was caused by oversupply and shifts in OPEC production plans.  Similar to this, the COVID-19 pandemic in 2020 caused significant market volatility because of a sharp decline in worldwide demand. The volatility in early 2022 was also influenced by the Russian-Ukrainian conflict, which represented how the market reacted to potential disruptions in supply.
Brent oil, a worldwide benchmark, is particularly sensitive to supply threats and OPEC decisions, as evidenced by its tendency to show more pronounced volatility increases during geopolitical events.  These results demonstrate how crucial realized volatility is to comprehend market behavior, and provides insights on how text analytics can be used to identify the main causes of price changes.

\subsection{Transforming News into Numerical Representations}

Sentiment analysis is an NLP technique used to assess the emotional tone of text, with applications in news, social media, and finance. Methods include lexicon-based approaches like VADER, which are fast but limited, and Transformer-based models like FinBERT, which offer higher accuracy in domain-specific contexts. Hybrid methods combine both to enhance performance across varied texts. For more details on how sentiment scores are computed for each one of those methods, refer to Appendix \ref{sec:appendix}.

To effectively utilize textual financial news for predicting oil price volatility movements, we additionally employ various LLMs to transform news articles into numerical representations (embeddings). Traditional models, such as GloVe and FastText, provide static word embeddings based on word co-occurrence statistics, while Transformer-based architectures like BERT and FinBERT generate contextualized embeddings that capture nuanced meanings in financial text. Hence, we will test both types.

We also include a domain-specific language model for financial applications called FinText.  In order to eliminate look-ahead bias and information leakage, FinText is trained on historical financial data using distinct models for various time periods. This ensures that embeddings do not integrate future knowledge.  For activities involving financial forecasting, this makes it very helpful.

We also use Google DeepMind's multimodal model Gemini and Meta's open-weight model LLaMA to investigate the potential of contemporary large-scale language models.  Despite being made for general-purpose language comprehension, these models are useful for a variety of NLP applications because of their sophisticated structures, which allow them to recognize intricate linguistic patterns.

We generate structured inputs appropriate for machine learning models by transforming news into vectorized embeddings.  We combine their embeddings by averaging to create a single daily feature vector, this is to normalize cases where there are several news articles available on a certain day.  As a result, market-related news is represented coherently and used as input for predictive models. For further details on the embedding process for each model and the aggregation technique applied, please refer to the repository \url{https://github.com/Romina-Hashami/Textual_Direction_Prediction_Oil_Volatility}.

\subsection{Textual Classification Framework}

This study develops a textual classification framework to predict the direction of crude oil market volatility using an ensemble learning approach. The framework consists of two key steps: (i) transforming textual data into numerical representations and (ii) applying a machine learning ensemble model with lagged features and a rolling window evaluation for classification.

To extract meaningful financial information from news text, we employ two separate feature extraction methods: embedding-based vectorization and sentiment analysis. These methods are applied independently to evaluate their individual effects on market movement predictions.

We utilize multiple large language models -- namely GloVe, FastText, BERT, FinBERT, Gemini, and LLaMA 3 -- to transform raw news text into numerical vectors. The vectorized representation of a given news article at time $t$ is denoted as:
\[ X_t = \psi(\text{text}_t) \]
where $X_t$ represents the extracted numerical feature vector, and $\psi$ is the embedding function of the selected LLM. Since multiple news articles are published per day, we aggregate the embeddings to ensure a single feature representation per day using the mean pooling method:
\[ X_{\text{daily},t} = \frac{1}{N_t} \sum_{i=1}^{N_t} X_{t,i} \]
where $N_t$ is the number of articles on day $t$, and $X_{t,i}$ is the vector representation of article $i$ on day $t$.

To assess the impact of news sentiment on volatility prediction, we extract sentiment scores using sentiment classifiers. The sentiment score for a given news text at time $t$ is defined as:
\[ S_t = \varphi(\text{text}_t) \]
where $S_t$ is the extracted sentiment score, and $\varphi$ represents the sentiment score function. Importantly, we do not combine sentiment scores with embedding-based features. Instead, we evaluate the predictive power of sentiment and embeddings separately in our classification framework.

We frame the volatility prediction problem as a binary classification task, where the target variable $Y_t$ is defined based on the realized volatility (RV):
\[ Y_t = \begin{cases} 
1, & \text{if } RV_t > RV_{t-1} \text{ (Volatility Increase)} \\
0, & \text{if } RV_t \leq RV_{t-1} \text{ (Volatility Decrease)}
\end{cases} \]

To enhance predictive accuracy, we introduce lagged features that incorporate past textual information. Instead of relying solely on daily aggregated features, we construct a feature matrix with multiple time-lagged versions of embeddings and sentiment scores up to $p$ days in the past:
\[ Z_t = \{X_t, X_{t-1}, \dots, X_{t-p}\} \text{ or } \{S_t, S_{t-1}, \dots, S_{t-p}\} \]

To classify volatility direction, we implement an ensemble learning approach, combining multiple classifiers in a Voting Classifier framework. The ensemble includes K-Nearest Neighbors (KNN), Naïve Bayes (NB), and Logistic Regression (LR). The final classification decision is determined through soft voting, where each model outputs a probability estimate, and the class with the highest aggregated probability is chosen:
\[ \hat{Y}_t = \arg\max_k \sum_{i=1}^n w_i P_i(Y_t = k) \]
where $P_i(Y_t=k)$ is the probability of class $k$ assigned by model $i$, and $w_i$ is the weight assigned to each model.

To ensure adaptability to market changes, we employ a rolling window evaluation, where the model is continuously updated over time. For training, we use 80\% of the data, ensuring that the model adapts to the most recent market conditions. The training and testing process follows these steps: At time $t$, the model is trained on the most recent 80\% of the available data. The trained model is then tested on the next single day ($t+1$) using all available news for that day. The training window is then rolled forward, and the process repeats. This rolling window approach prevents the model from being biased toward outdated historical patterns and ensures that it learns from the most relevant market conditions.

\subsection{Interpreting the Model}
In this section, we explain how we interpret the predictions of our classification model, using the SHAP framework to understand the contributions of individual words to the model’s output. SHAP, as discussed in \cite{shapley1953value, lundberg2017unified}, originates from cooperative game theory and is widely used in machine learning to fairly distribute the contribution of each feature towards the model’s predictions. We can learn more about which textual elements have the biggest impact on the model's decision-making process by calculating the impact of individual words, phrases, or entire articles on the categorization result.

We can examine how each word contributes to the model's prediction by using SHAP.  This enables us to draw attention to the precise words that the model uses most frequently, providing us with information about how the model processes news. Since our features are numerical embeddings, each SHAP value reflects the influence of a specific embedding dimension on the model's prediction of next-day realized volatility (RV). To interpret these values in terms of actual words, we tokenized the original news texts (see Table \ref{tab:token_outputs_clean}) and estimated the impact of each word by distributing the total SHAP contribution across all meaningful words in each article. This allowed us to approximate the average importance of individual words and identify which terms consistently influenced the model’s decisions.

This kind of analysis is especially important when working with large volumes of textual data. Instead of treating news articles as a single block of information, SHAP allows us to see the individual impact of each word, giving us a much more detailed understanding of how the model interprets daily news.

In our approach, instead of averaging high-dimensional textual vectors, which could lead to excessive smoothing of information and the potential dilution of valuable signals, we use all the available data for a given day. This means that each individual news article’s impact is preserved and used to interpret the model’s classification. In particular, we use the day's data as the model input, enabling SHAP to assess the contribution of each individual article to the volatility direction prediction at the end.  This approach guarantees that no news item is overlooked and that every article's complete impact is recorded for the interpretation process.

For one-day-ahead forecasting, we use the news data from the current day to predict the volatility direction for the next day. The classification model then processes this input, and the SHAP framework allows us to interpret the contributions of the individual news articles in determining the predicted volatility direction for the following day.

In accordance with \cite{xu2024volatility}, we divide the dataset into four discrete periods that correspond to significant world events like the COVID-19 pandemic and the Russia-Ukraine conflict. These time frames depict how the oil market changed during these major shocks. For each period, we present the top words with the highest average SHAP impact, revealing which terms were most influential in shaping the model’s predictions.

\subsection{McNemar Test}

The McNemar test is a non-parametric statistical method that compares two related classification models or treatments on the same dataset. Introduced by McNemar \cite{mcnemar1947note}, it is beneficial for paired categorical data, such as binary classification outcomes. Unlike standard accuracy comparisons, the test examines cases where the two models disagree, providing insight into whether one method consistently outperforms the other. In this study, since different feature sets are used within the same classification model, the McNemar test is an appropriate tool to assess whether the observed differences in predictive performance are statistically significant.

\section{Results}

The results of our rolling window approach, using 80\% of the data for training and 20\% for out-of-sample validation, provide valuable insights into the predictive performance of different sentiment analysis and embedding methods for forecasting realized oil volatility. The HAR  model is a recognized benchmark for volatility forecasting, which is used as a baseline for comparison. This model is suited for predicting realized volatility, as it captures the time-dependent nature of volatility by modeling long memory in financial time series. By comparing news-based models to HAR, we can assess the added value of incorporating sentiment analysis and advanced embedding methods into volatility forecasting.  

For the HAR model, we considered the previous day's volatility, last week's, and last month's features. However, in sentiment-based methods, only sentiment-related features were used, and in embedding-based methods, only the embedding vectors were considered as features. This design allows us to analyze whether these features alone can improve prediction performance without additional help from historical volatility data.  

Table \ref{tab:sentiment_results} reports the performance metrics for sentiment-based models, including accuracy, precision, recall, and F1-score, which provide a comprehensive evaluation of model performance.

The HAR model alone achieved an accuracy of 0.6494, serving as a baseline. Among sentiment-based models, the Count method outperformed others with an accuracy of 0.7054, which is also statistically significantly better than the HAR model based on the McNemar test result. This improvement indicates that simply counting news articles (linked to higher volume of news articles) contributes to volatility forecasting. In contrast, VADER, TextBlob, FinBERT, and CrudeBERT performed worse than HAR, highlighting that traditional sentiment scoring may not be sufficient for this task, and why we further study embedding methods. 

\begin{table}[h]
\caption{Performance metrics for sentiment-based models. In bold: best performance. Underlined: second best performance.}
\vspace{1em}  
\label{tab:sentiment_results}
\centering
\begin{tabular}{l@{\hspace{8mm}}cccc} 
\toprule
Model & Accuracy & Precision & Recall & F1-Score \\
\midrule
HAR & \underline{0.6494} & \underline{0.6491} & \underline{0.6494} & \underline{0.6488} \\
vader & 0.5777 & 0.5818 & 0.5777 & 0.5770 \\
textblob & 0.5777 & 0.5766 & 0.5777 & 0.5749 \\
finbert & 0.5368 & 0.5354 & 0.5368 & 0.5352 \\
crudebert & 0.6105 & 0.6125 & 0.6105 & 0.6032 \\
count & \textbf{0.7054} & \textbf{0.7065} & \textbf{0.7054} & \textbf{0.7037} \\
\bottomrule
\end{tabular}
\end{table}

Table \ref{tab:embedding_results} presents the performance of embedding-based models. FastText achieved the highest accuracy (0.7136), outperforming all other embedding methods. Llama, Gemini, BERT, FinBERT, and GloVe also improved over HAR, whereas only FinText performed worse. The results suggest that FastText's ability to handle subword information (see Appendix \ref{sec:appendix}) contributes significantly to a task such as volatility prediction. 

\begin{table}[h]
\caption{Performance metrics for embedding-based models. In bold: best performance. Underlined: second best performance.}
\vspace{1em}  
\label{tab:embedding_results}
\centering
\begin{tabular}{l@{\hspace{8mm}}cccc} 
\toprule
Model & Accuracy & Precision & Recall & F1-Score \\
\midrule
HAR & 0.6494 & 0.6491 & 0.6494 & 0.6488 \\
GloVe & 0.6661 & 0.6714 & 0.6661 & 0.6602 \\
Fasttext & \textbf{0.7136} & \textbf{0.7189} & \textbf{0.7136} & \textbf{0.7098} \\
bert & 0.6743 & 0.6786 & 0.6743 & 0.6695 \\
finbert & 0.6694 & 0.6811 & 0.6694 & 0.6597 \\
fintext & 0.5925 & 0.5916 & 0.5925 & 0.5905 \\
gemini & 0.6858 & 0.6909 & 0.6858 & 0.6810 \\
llama & \underline{0.6890} & \underline{0.6888} & \underline{0.6890} & \underline{0.6886} \\
\bottomrule
\end{tabular}
\end{table}


To further analyze these results, we conducted McNemar’s test to determine statistical significance. The detailed results are presented in the Appendix: Table \ref{tab:mcnemar_sentiment} shows the McNemar test results for sentiment-based models, while Table \ref{tab:mcnemar_embedding} presents the results for embedding-based models.

The McNemar test confirms that the Count model significantly outperforms the HAR model (p = 0.0392), underline that counting the number of news articles can statistically enhance prediction accuracy. Furthermore, among embedding-based models, FastText was the only model that significantly outperformed HAR (p = 0.01277), whereas other embeddings did not achieve statistical significance despite performing better in evaluation metrics. 

The results indicate that sentiment-based and embedding-based models can enhance volatility forecasting. Still, only certain approaches, such as FastText and Count-based sentiment analysis, show statistically significant improvements. The findings highlight the importance of selecting appropriate text-based features and embedding strategies when integrating sentiment analysis into financial forecasting models.

\subsection{Interpreting the Model}
In this section, we present the results of our model interpretability analysis, which leverages the SHAP framework to understand the contributions of individual words to the model's predictions on realized volatility. In the SHAP framework, the values represent the contribution of each feature to the model's predicted output (each one of them between 0 and 1, with 0 the lowest contribution, and 1 the highest). SHAP values quantify how much each feature alters the model's prediction from the baseline, which is the model's output when no features are considered, indicating how much a specific feature impacts the prediction relative to this baseline.
Since the model uses word embeddings as features, word-level analysis is made easier by mapping the SHAP values for these embeddings back to specific words in the text.  This makes it possible to pinpoint the precise terms that most significantly affect the model's predictions. In these figures, the x-axis represents the SHAP values, which indicate the importance of each word, and the y-axis lists the corresponding words.

\begin{figure}[h]
\centering
\includegraphics[width=7.5cm]{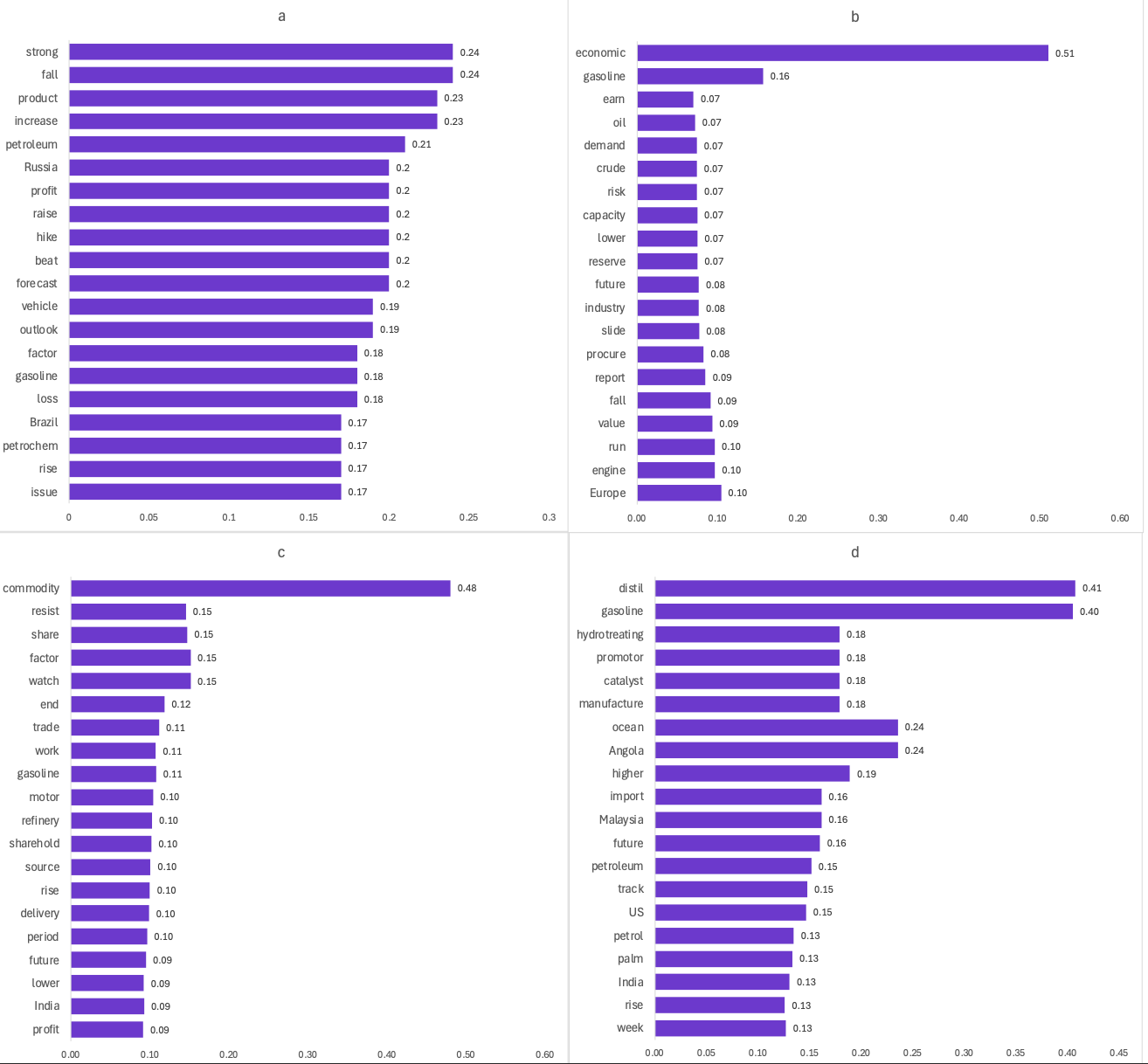}
\caption{Top 20 Most Influential Words by SHAP Impact Across Four Time Periods:  
    (a) Before the Coronavirus Outbreak, (b) Epidemic Shock Period, (c) Epidemic Stabilization Period, (d) Russia-Ukraine Conflict.}
    \label{fig:intrepret}
\end{figure}

In the pre-pandemic period (Figure~\ref{fig:intrepret}a), the model placed significant importance on terms like \textit{issue, rise, gasoline, and petroleum}. Words associated with economic indicators (\textit{forecast, factor, product}) and market movements (\textit{increase, fall}) also contributed notably to predictions of volatility direction. This suggests that market participants in this period were primarily concerned with supply-demand factors and market expectations regarding economic growth, rather than external shocks.

During the early stages of the pandemic (Figure~\ref{fig:intrepret}b), the model's most influential words included {\it economic} (with a particularly high SHAP value of 0.51), reflecting the heightened focus on economic instability and uncertainty. The global scope of the disruption and the particular industries, such as energy, that were most vulnerable to the shock are highlighted by other significant terms like {\it gasoline, Europe,} and {\it crude}. This change in word importance is consistent with the pandemic's crisis-driven instability.

In the post-shock stabilization period (Figure~\ref{fig:intrepret}c), the most influential words were those connected to broader market trends and economic factors. For example, terms such as \textit{commodity, watch, and resist} suggest that market participants were paying attention to the long-term impacts of the pandemic and the broader global economy. Words like \textit{gasoline and trade} indicate the importance of oil demand and supply chain disruptions during this phase.

During the Russia-Ukraine war (Figure~\ref{fig:intrepret}d), terms like \textit{import, fuel, and distill} were found to have a significant impact. The increased geopolitical risk and the resulting disruptions in energy markets are reflected in these phrases. Terms like \textit{Angola, Malaysia, and petroleum} are used to emphasize how the conflict has changed the systems of oil production and distribution. The model also shows an increased sensitivity to terms related to specific oil-producing countries and oil products, indicating the regional impacts of the conflict on global markets.
\vspace{1.5em}

By breaking down the analysis into these four periods, we observe how the model's attention shifts in response to major global events. SHAP values allow us to understand the dynamic nature of the market's reaction to external shocks and the way in which textual data, particularly news articles, informs predictions of volatility. In this context, the SHAP framework not only provides a way to interpret the model's decisions but also sheds light on the evolving factors driving market behavior over time.

This analysis shows which factors moved the markets at different times by examining how the model's forecasts were affected by events like the pandemic and international conflicts.  It helps us better understand the relationship between news and market behavior, which may be helpful for enhancing future projections and enabling us to make more informed judgments when the market becomes volatile.

\section{Conclusion}

 This study uses ensemble learning approach combining Logistic Regression, Naïve Bayes, and K-Nearest Neighbors to see the potential of news data in improving oil price volatility direction forecasting. The findings highlight the value of sentiment-based and embedding-based approaches in enhancing predictive accuracy. Notably, the Count-based sentiment model and the FastText embedding method emerged as the most effective strategies, significantly outperforming the benchmark Heterogeneous Autoregressive (HAR) model in predicting volatility direction. The McNemar test shows that the model using number of news per day, achieved the p\text{-value} of 0.0392 which  is a statistically significant improvement over all sentiment methods and HAR model, suggesting that the volume of news coverage alone provides meaningful insights into market behavior. Similarly, the superior performance of FastText (p = 0.01277) reflects the importance of subword-level information in capturing the complexities of financial text data.
 
Despite the greater complexity of other embedding models, the relatively simple FastText method outperformed them all. This suggests that the ability to handle subword information and morphological variations plays a more important role in forecasting accuracy than model complexity itself. 
These predictive improvements were achieved without incorporating past volatility data. The sentiment-based and embedding-based models in this study exclusively used news-derived characteristics, in contrast to the HAR model, which is based on historical volatility patterns.  This challenges the conventional dependence on historical volatility data in financial forecasting by showing that textual data alone can produce useful predictive signals.
This study contributes to the growing body of research on the integration of textual input into financial models by offering helpful insights on the selection of sentiment and embedding strategies.   Future research should look into hybrid modeling frameworks and other text processing methods to further increase predicted accuracy. The shown performance of the Count and FastText models emphasizes the importance of appropriately selecting and developing text-based features by showcasing the value of news data in enhancing market volatility forecasts.

Future work could investigate modeling the sequence of daily news-derived features in a time series framework, allowing the prediction of volatility to depend on the temporal dynamics of textual information rather than treating each day independently. In order to better capture the most informative elements of daily news information, various embedding aggregation algorithms beyond mean pooling, such as weighted pooling, max pooling, or attention-based aggregation, should be investigated. This is because several news items occur every day.





\bibliographystyle{unsrtnat} 
\bibliography{mybibfile}
\clearpage

\appendix
\section*{Appendix}
\section{Sentiments and Embeddings models}\label{sec:appendix}

In this section, we provide a description of the sentiment and embeddings methods used in this study

\begin{description}
    \item[TextBlob] is a Python library designed for sentiment analysis, providing polarity and subjectivity scores based on a predefined lexicon. It processes textual data at the sentence level, effectively handling negations to determine sentiment on a scale from -1  to 1. 
    \vspace{.7em}

    \item[VADER (Valence Aware Dictionary and sEntiment Reasoner)] is a rule-based sentiment analysis tool optimized for social media and short texts. It assigns sentiment scores based on a lexicon of human-annotated words, capturing both intensity and polarity. VADER is particularly effective at recognizing informal language, punctuation, and emojis, making it a fast and efficient tool for sentiment classification.      \vspace{.7em}

\item[BERT (Bidirectional Encoder Representations from Transformers)] is a language model introduced by Devlin et al. (2018) \cite{devlin2018bert} that learns word embeddings using a bidirectional Transformer architecture. Unlike Word2Vec and GloVe, which produce static word representations, BERT captures contextual meaning by processing words in both left and right directions. 
BERT is trained on BooksCorpus (800M words) and English Wikipedia (2,500M words) using masked language modeling (MLM) and next sentence prediction (NSP). 
\vspace{.7em}

\item[FinBERT] is a specialized adaptation of BERT, designed for financial text analysis. Introduced by Araci (2019) \cite{araci2019finbert}. FinBERT is pretrained on financial texts, such as analyst reports, earnings calls, and financial news, improving its ability to interpret domain-specific terminology and sentiment. FinBERT employs masked language modeling and fine-tuning techniques on datasets like the Financial PhraseBank.
\vspace{.7em}

\item[CrudeBERT] is a sentiment analysis model fine-tuned from FinBERT for the crude oil market. It leverages domain-specific datasets to better capture sentiment dynamics related to oil price movements. By incorporating supply and demand factors, CrudeBERT improves sentiment classification for market-relevant news, aiding in more accurate forecasting of crude oil price trends. 
\end{description}

\begin{description}
    \item[GloVe (Global Vectors for Word Representation)] is a method for generating word embeddings. The model is trained over billions of tokens of data, including Wikipedia dumps and Common Crawl, to capture global word relationships through co-occurrence statistics \cite{pennington2014glove}. Pretrained GloVe embeddings are available in various sizes, and can be freely accessed on platforms like GitHub.  For our study, we use the 100-dimensional pretrained GloVe model to generate embeddings and capture the semantic context of financial news articles.

\vspace{.7em}

\item[FastText] introduced by the Facebook AI Research team in \cite{bojanowski2017enriching}, is a word embedding model designed to improve upon traditional models like Word2Vec. Unlike Word2Vec, which treats each word as an atomic unit, FastText represents words as sequences of character n-grams. FastText may record subword information using this technique, which is very helpful when dealing with uncommon or non-vocabulary terms. FastText can produce more accurate and reliable embeddings by dissecting words into smaller, meaningful subword components (such as prefixes and suffixes).
\vspace{.7em}

\item[FinText] is a specialized financial language model introduced in \cite{rahimikia2024r}.  FinText consists of historical, year-specific language models trained on financial and accounting texts from 2007 to 2023, effectively mitigating look-ahead bias, which guarantees that it can only access data that is available at the moment.  
\vspace{.7em}

\item[LLaMA (Large Language Model Meta AI)] is a family of Transformer-based language models developed by Meta AI, designed to provide high-quality text representations and generation capabilities. LLaMA employs a causal decoder-only Transformer architecture, allowing it to capture rich contextual meanings and generate embeddings that reflect both semantic and syntactic nuances \cite{touvron2023llama}.
LLaMA operates as an autoregressive model, predicting the next token in a sequence. This makes LLaMA particularly effective for natural text generation and document embeddings. In our work, we leverage LLaMA 3 to transform textual data into dense vector representations. Specifically, we extract embeddings from the model's hidden states.
\vspace{.7em}

\item[Gemini]  family of multimodal models developed by Google DeepMind, designed to handle and integrate multiple data types, including text, images, audio, and video \cite{anil2023gemini}. For text vectorization, Gemini leverages its deep Transformer-based architecture to generate high-quality contextual embeddings. The model processes input text through tokenization and normalization, after which embeddings are extracted from its hidden states \cite{anil2023gemini}. In particular, fixed-length numerical representations of text are generated using mean-pooled embeddings from the last Transformer layer.  These embeddings are ideal for downstream tasks like document classification, clustering, and semantic search because they capture both syntactic and semantic links.

\end{description}
\section{McNemar’s Test Results}

\begin{table}[H]
\caption{McNemar test results for sentiment-based models.}
\vspace{0.5cm}
\label{tab:mcnemar_sentiment}
\centering
\begin{tabular}{l@{\hspace{4mm}}lllll}
\toprule
 & Count & HAR & CrudeBERT & VADER & TextBlob \\
\midrule
HAR & 0.0392 & - & - & - & - \\
CrudeBERT & 0.0002 & 0.1442 & - & - & - \\
VADER & 5.7E-07 & 0.0066 & 0.241 & - & - \\
TextBlob & 9.6E-07 & 0.0176 & 0.3504 & 0.803 & - \\
FinBERT & 1.1E-09 & 4.1E-05 & 0.0079 & 0.16 & 0.0901 \\
\bottomrule
\end{tabular}
\end{table}

\begin{table}[H]
\caption{McNemar test results for embedding-based models.}
\vspace{0.5cm}
\label{tab:mcnemar_embedding}
\centering
\begin{tabular}{l@{\hspace{1mm}}l@{\hspace{3mm}}l@{\hspace{3mm}}l@{\hspace{3mm}}l@{\hspace{3mm}}l@{\hspace{3mm}}l@{\hspace{3mm}}l}

\toprule
 & FastText & Llama & Gemini & BERT & FinBERT & GloVe & HAR \\
\midrule
Llama & 0.2566 & - & - & - & - & - & - \\
Gemini & 0.0667 & 0.8221 & - & - & - & - & - \\
BERT & 0.0153 & 0.4017 & 0.6022 & - & - & - & - \\
FinBERT & 0.0104 & 0.3271 & 0.4118 & 0.8357 & - & - & - \\
GloVe & 0.0032 & 0.2461 & 0.3533 & 0.7545 & 1.0000 & - & - \\
HAR & 0.0127 & 0.1540 & 0.2260 & 0.4258 & 0.5384 & 0.5782 & - \\
FinText & 1.7E-07 & 8.6E-05 & 0.0002 & 0.0006 & 0.0027 & 0.0019 & 0.0562 \\
\bottomrule
\end{tabular}
\end{table}
\end{document}